\documentclass[twocolumn]{aastex61}
\usepackage{natbib}
\usepackage{amsmath,bm}
\usepackage{subfigure}

\begin{document}

\title{Flows around Averaged Solar Active Regions}

\author{D.~C.~Braun}
\affiliation{NorthWest Research Associates, 3380 Mitchell Lane, Boulder, CO 80301, USA}
\email{dbraun@nwra.com}

\begin{abstract}
We explore the general properties of near-surface                                                   
flows around solar active regions.                                                                
Helioseismic holography is applied to HMI Dopplergrams yielding 
nearly 5000 flow measurements of 336 unique 
active regions observed by the Solar                    
Dynamics Observatory between 2010 and 2014.                                                         
Ensemble averages of the flows, over subsets of regions sorted on the basis of magnetic             
flux, are performed.  These averages show that 
converging flows, with speeds                             
about 10 m s$^{-1}$ and extending up to 10$^{\circ}$ from the                                 
active region centers, are prevalent and have similar properties for all regions                    
with magnetic flux above $10^{21}$ Mx. Retrograde flows are also detected, with amplitudes          
around 10 m s$^{-1}$, which predominantly, but not exclusively, flank the polar side of the 
active regions.
We estimate the expected contribution of these active-region flows to longitudinal                  
averages of zonal and meridional flows and demonstrate the plausibility that                        
they are responsible for at least some component of the time-varying global-scale 
flows.                  
The reliability of our flow determination is tested using publicly available 
MHD simulations of both quiet-Sun convection and of a sunspot. 
While validating the overall methodology in general, the sunspot simulation
demonstrates the presence of artifacts which may compromise quantitative flow
inferences from some helioseismic measurements.
\end{abstract}

\section{Introduction}

The detection of flows with amplitudes of order 10 m s$^{-1}$ or more and 
converging towards active regions (hereafter ARs) was an early discovery
in local helioseismology \citep{Gizon2001,Haber2004,Zhao2004}. 
However, their general characteristics, including their detailed variation with 
magnetic properties of
the associated magnetic regions, remain largely uncharted. These flows appear to provide a
link between convection and magnetism, two critical processes governing the solar convection zone.
Possibly lasting (at least) as long as the magnetic regions themselves, the
flows are suspected of affecting larger circulation patterns, particularly the meridional
and zonal flow components of global circulation. For example, converging flows in active
region latitudes appear to modulate and reduce the 
amplitude of the meridional flow pattern \citep{Chou2001,Gizon2003,Zhao2004}. The
meridional flow is involved critically in the process that leads to 
polar field polarity reversals \citep{Wang2002} and the ability to predict properties
of subsequent solar cycles \citep[e.g.\ as reviewed by][]{Jiang2014}. 
The modulation associated with active latitudes,
and which may include AR-related inflows, is believed
to provide critical nonlinear feedback, preventing eventual decay or growth of
subsequent solar cycles \citep[e.g.][]{Jiang2010,Cameron2012,Martin-Belda2017}.
The role of this
modulation has been explored as a contributor to the recent extended solar minimum
and weak cycle 24 \citep{Upton2014}.

Early measurements were made of the subsurface flows beneath individual active regions 
\citep{Zhao2004,Haber2004}, which are described as a toroidal-like circulation
with converging flows extending down to depths of about 10 Mm and diverging flows below this
depth. As reviewed by \cite{Gizon2010b} this scenario is far from established and 
relevant systematic surveys have been sparse.  
Some general properties of the flows and their relation to other AR properties have
been studied \citep{Komm2007,Komm2011c,Komm2015c}, albeit with helioseismic methods with
relatively low spatial resolution (i.e. having spatial scales on the order of 15$^{\circ}$,
which is comparable to the AR sizes).  The ring-diagram based
survey performed by \cite{Hindman2009} focused on the
near-surface ($< 2$ Mm deep) flow properties, including inflow and circulation speeds, 
of $\sim$ 100 active regions at somewhat higher spatial resolution
($\sim 2^{\circ}$) .

Studies of near-surface flows using
helioseismic methods with greater spatial resolution are hampered by the 
strong flows associated with supergranulation. Supergranules 
with peak flows $\sim$ 300 m s$^{-1}$ \citep{Rincon2018}, and root-mean-squared fluctuations 
$\sim 100$ m s$^{-1}$, effectively act as noise and 
dominate the weaker flows associated with ARs. 
Ensemble averaging, consisting of identifying and averaging coaligned
flow measurements of features with (expected) similar properties provides one method
for increasing the signal-to-noise ratio. \cite{Loeptien2017}, using a local-correlation
tracking method applied to the granulation pattern, carried out such averaging over
more than 200 active regions.
In this work, we carry out a high-spatial
resolution ($\sim 1^{\circ}$) survey of active regions flows, using helioseismic holography 
(hereafter HH) applied to Dopplergrams obtained from the Helioseismic and Magnetic Imager (HMI).
Our intent is to measure and compare ensemble-averaged AR flows across a 
wide range of magnetic flux.  To accomplish this, we use both existing 
\citep{Braun2016} flow measurements of 252 large (NOAA numbered) ARs,
as well as additional measurements of flows around regions with   
magnetic fluxes as low as $10^{21}$ Mx. Our survey is discussed in 
\S\ref{sec.survey}, which includes a description of AR identification 
(\S\ref{sec.select}), and the magnetic properties of the complete sample 
(\S\ref{sec.magnet}).  The ``calibrated helioseismic holography'' method, 
employed to infer the near-surface flows, is described in \S\ref{sec.ttmeas}. 
Results are presented in \S\ref{sec.results}, followed by a discussion in
\S\ref{sec.discuss}.

\section{Survey}\label{sec.survey}

The starting point of our flow survey is the helioseismic 
holography analysis by \cite{Braun2016}. 
This prior survey, carried out to probe the relationship
between flows and solar flares, produced approximately 4000 sets of near-surface flow maps 
of 252 unique NOAA numbered sunspot regions present between 2010 May and 2014 
December. The use of the largest sunspot groups (as ranked by sunspot areas) 
was appropriate for studies of solar flares,
but for the present work we have extended the AR sample  
to include weaker regions. The method used to identify these additional regions is 
described below.  

\subsection{Selection}\label{sec.select}

To achieve a representative sample of ARs for a given solar rotation we start with 
HMI synoptic magnetograms. Taking the absolute value of the magnetic flux density, we 
applied spatial smoothing with a two-dimensional Gaussian function with a full-width 
at half-maxima (FWHM) of 10$^{\circ}$.  We located all of the peaks in 
this smoothed map, where a peak is defined to have a pixel value greater than the
eight neighboring pixels.  Sorting the peaks from highest to lowest magnetic flux density, 
we discard those which are situated within 20$^{\circ}$ of any larger peak. 
The total unsigned flux of each candidate AR (contained within a 
20$^{\circ}$ $\times$ 10$^{\circ}$ bounding box) were assessed from the synoptic magnetogram, 
and only regions with a total flux greater than $10^{21}$ Mx (the lower limit 
of this survey) were retained.  While the method 
is not intended to identify all possible magnetic regions, it does select ones which 
are more spatially separated from each other. This allows the flows associated with those regions 
to be more readily isolated. 

\begin{figure*}
\plotone{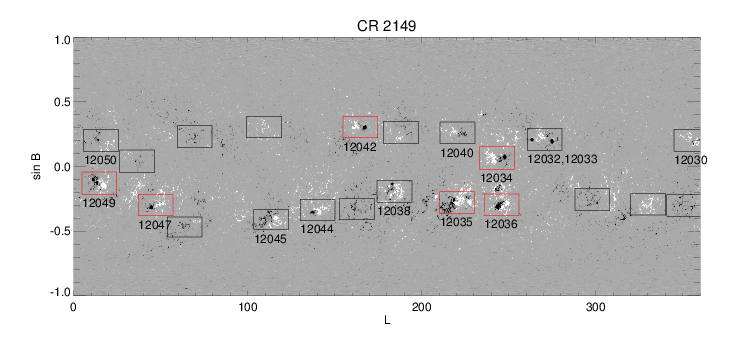}
\caption{
SDO/HMI synoptic magnetogram for Carrington rotation 2149, showing some active
regions selected for this survey.
Red boxes enclose large sunspot groups which were previously
identified and analyzed \citep{Braun2016} on the basis of NOAA published sunspot areas.
Black boxes indicate (mostly weaker) regions which are included in this survey using the
identification method described in the text.
Some of these regions are labeled by one (or more) NOAA sunspot numbers; 
others have no NOAA designation.
}
\label{fig.cr2149}
\end{figure*}

We employ this procedure to six Carrington rotations, spaced ten rotations apart, and
spanning the time range of the \cite{Braun2016} survey (specifically, these
included Carrington rotation numbers CR 2099, 2109, 2119, 2129, 2139, and 2149). 
An example of the regions identified for CR 2149 (the most active rotation) 
is given by Figure~\ref{fig.cr2149}.
A total of 104 ARs are identified in this manner during the six rotations, of which 20 
are part of the \cite{Braun2016} survey.
Our complete sample thus contains 336 unique ARs: 252 from the original flare survey plus
84 new regions.

\subsection{Magnetic Properties}\label{sec.magnet}

Datacubes of Dopplergrams and magnetograms, centered on the AR locations determined
using the procedure described above, are constructed identically to those employed
by \cite{Braun2016}. 
Specifically, full-disk HMI magnetograms and Dopplergrams are remapped to
Postel coordinates ($x_p$, $y_p$) with a tangent point centered on the AR 
location, tracked with a fixed Carrington rate, and spanning 
30$^{\circ}$ by 30$^{\circ}$ with a pixel spacing of 0.0573$^{\circ}$. 
The choice of the Carrington rate, historically derived from observations of
sunspots and defined as one rotation per 27.2753 days as viewed from Earth,
is motivated by the desire to minimize spatial drifting of ARs in the Doppler and magnetogram
time series.  Further analysis to remove contributions from large-scale flows, including departures
from the Carrington rotation rate, is described in \S\ref{sec.detrend}.
We use full-disk Dopplergrams with the full cadence of
45 seconds for the HH analysis of flows discussed in \S\ref{sec.ttmeas}. 

The magnetic properties of each AR are studied from remapped full-disk magnetograms
sampled every 68 minutes.  The passage of each AR across the disk is divided into
16 non-overlapping intervals each spanning 13.6 hr.
To ensure the quality of the helioseismic analysis,
intervals for which the AR position was farther than 60$^{\circ}$
from disk center, or for which gaps in the HMI Dopplergram data exceeded 30\% of the 13.6 hr period
are excluded from further study. 

For each AR, the Postel-remapped magnetograms are averaged over each of the 13.6 hr intervals.
For the purpose of coalignment necessary for ensemble averaging, 
positions defining the center of mass (centroid) of the unsigned magnetic flux density, relative
to the Postel-projection center, are obtained. During this step, the net unsigned flux (corrected
for the cosine of the heliocentric angle) is also integrated and recorded. 
To reduce the effect of magnetogram noise, only pixels with
flux density greater than 50 G present in a 20$^{\circ}$ $\times$ 10$^{\circ}$ bounding box are retained in 
the centroid and flux determinations.  The statistics of these position offsets were examined,
and a rejection of time intervals for which the offset (in either the $x_p$ or $y_p$ 
coordinate of the 
Postel frame) exceeded the mean by plus-or-minus three standard-deviations was carried out.  
Values of the standard deviations are 1.2$^{\circ}$ and 0.6$^{\circ}$ in the $x_p$ and $y_p$ 
directions, respectively. 

\begin{figure}
\plotone{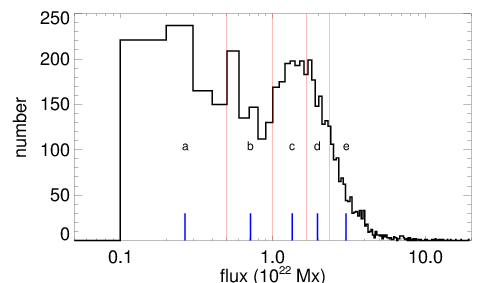}
\caption{
Histogram of the magnetic flux determined over 4925 time intervals of 
magnetograms of the 336 active regions studied in this work. Red lines delineate
the flux groups selected for study, with the blue markers indicating the median
flux of each group sample.
}
\label{fig.hist}
\end{figure}

\begin{figure*}
\plotone{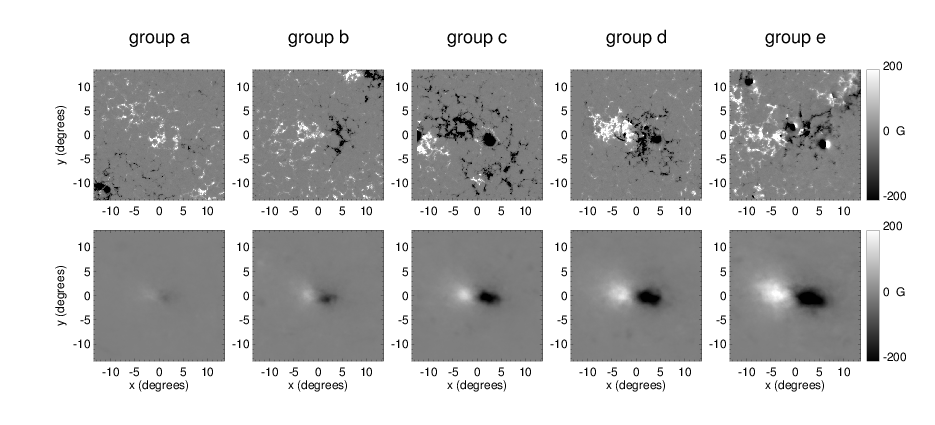}
\caption{
Magnetic appearance of active regions divided into five groups 
of increasing total magnetic flux.
The top row 
shows 13.6-hr averages of magnetograms for a 
randomly selected AR sample in each group while the column below that shows the ensemble average
of the signed magnetic flux density after coaligning the
magnetograms to their center of mass (as determined from the unsigned flux density). 
The coordinate system is defined as the distance from the center-of-mass towards the 
west ($x$) and towards the pole ($y$) of the hemisphere in which the AR resides. 
}
\label{fig.mags}
\end{figure*}

After the analysis and data rejection described above, we are left with 4925 
measurement sets which exceed $10^{21}$ Mx flux.  A histogram of the fluxes
is shown in Figure~\ref{fig.hist}. The distribution is divided into five flux
groups as separated by the vertical lines in the figure. Some properties of the groups are
listed in Table~\ref{tbl.flux}, including the defining flux range, the median flux 
(${\Phi}_\mathrm{med}$) for each group, 
and the number of measurements within each group. 
The magnetic flux in most ARs change over the 9 days they are tracked, 
resulting in most regions being included in more than a single flux group. 
Magnetograms of ARs, randomly selected from each group, are shown in the top row of 
Figure~\ref{fig.mags}. 
Coaligned averages over each group of the signed line-of-sight magnetic flux density
are shown in the bottom row of Figure~\ref{fig.mags}.
For the averaging, we adapt a coordinate system ($x$,$y$) with an origin at the center-of-mass, $x$ 
increasing in the westward (prograde) direction, and
$y$ increasing towards the pole of the hemisphere
in which the AR resides. Thus, ARs in the southern hemisphere are spatially 
flipped around the $x_p$ axis.
In accordance with Hale's polarity law, regions in the southern
hemisphere also have their polarity switched. 
The group averages clearly show the
two polarities, tilted by Joy's law, and having an asymmetry in peak flux density 
between polarities.

\begin{table}[ht!]
 \begin{center}
  \caption{properties of magnetic flux groups}\label{tbl.flux}
  \begin{tabular}{cccc}
   \tablewidth{0pt}
   \hline
   \hline
   {group} & ${\Phi}$ range & ${\Phi}_\mathrm{med}$ &  number \\
      {} &  ($10^{22}$ Mx)   & ($10^{22}$ Mx)    &    {} \\
   \hline
      a  &       0.1 -- 0.5   &       0.27        &  773   \\
      b  &       0.5 -- 1.0   &       0.71        &  733   \\
      c  &       1.0 -- 1.67  &       1.34        &  1234  \\
      d  &       1.67 -- 2.35 &       1.97        &  1088  \\
      e  &       $ > 2.35$   &       3.01        &  1097  
  \end{tabular}
 \end{center}
\end{table}

\subsection{Calibrated Helioseismic-holography Flows}\label{sec.ttmeas}

\begin{figure}
\plotone{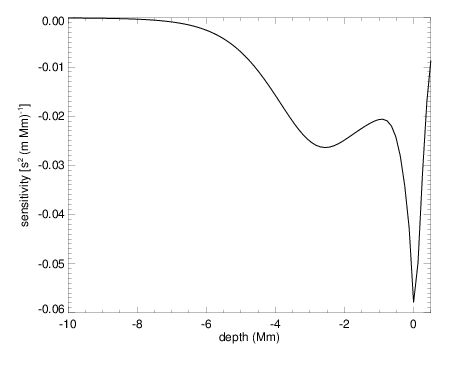}
\caption{
Depth dependence of the horizontally integrated sensitivity kernel, relating the contribution
of a horizontal
flow component to the lateral-vantage travel-time difference as measured across opposite 
quadrants and for a nominal focus depth of 3 Mm. This function roughly describes the
relative weight in depth of the true flows contributing to our calibrated flow maps.
There is a  peak in the sensitivity at the photosphere and a weaker peak around
the nominal focus depth.  
Adapted from Figure 13 of \cite{DeGrave2018}.
}
\label{fig.kernz}
\end{figure}

We use helioseismic holography in the so-called lateral-vantage geometry 
\citep[e.g.][]{Lindsey2004b} and with a focus depth of 3 Mm below the photosphere. 
This method is analogous to deep-focus
methods in time-distance helioseismology and common-depth-point reflection
terrestrial seismology. The result is the determination of travel-time differences
of waves propagating between opposite quadrants of an annular pupil. As was carried out in
our prior analysis \citep{Braun2016}, we use a simple numerical calibration to directly
convert the travel-time differences to components of the flow. This calibration is
based on assumptions which include: 1)
the sensitivity function, which relates the three-dimensional flow properties to the 
resulting travel time difference, is compact in volume relative to the spatial scale of
the inferred flows, and 2) the horizontal flow components (rather than vertical ones) 
produce the principle contribution to the time differences. 
Similar calibration procedures have been extensively used, particularly
for the study of shallow flows, with both $f$-modes \citep{Gizon2001,Gizon2003b} and
high-degree $p$-modes \citep{Braun2004,Braun2011a,Birch2016,Braun2016}. The calibration
constant, relating the west-minus-east (WE) and north-minus-south (NS) travel-time
difference into westward and northward vector components is deduced by the application
of two different tracking rates to a time-series of Dopplergrams of a region on the Sun.
The weighting in depth of the subsurface horizontal flows which contribute to the measurements
can be (roughly) characterized by the horizontal integral of the sensitivity function.
For the measurements used here, this weighting is plotted in Figure~\ref{fig.kernz}. The sensitivity
function shown is derived using the Born approximation and is described
by \cite{DeGrave2018}.

Comparisons between results obtained from calibrated-HH methods
and non-helioseismic procedures (e.g.\ local
correlation tracking) have been successfully performed \citep[e.g.][]{Birch2016}.
The construction of realistic numerical simulations of wave propagation within
solar-like model interiors has allowed the direct testing of helioseismic methods 
\citep[e.g.][]{Braun2007,Zhao2007,Braun2012,DeGrave2014b,DeGrave2018} 
In Appendix~\ref{sec.valid} we use publicly 
available MHD simulations, of both solar convection and realistic sunspots, 
to validate the calibrated flows used in this work.

\section{Results}\label{sec.results}

Ensemble averages of the calibrated-HH flow components for each flux group
are made, after coaligning each flow map. This alignment uses the same
center-of-mass positions determined from the magnetograms (\S\ref{sec.magnet}).
The remapping to the center-of-mass distances ($x$,$y$) is performed on 
each flow component using bicubic interpolation. 

\subsection{Large-scale Detrending}\label{sec.detrend}

\begin{figure*}
\plotone{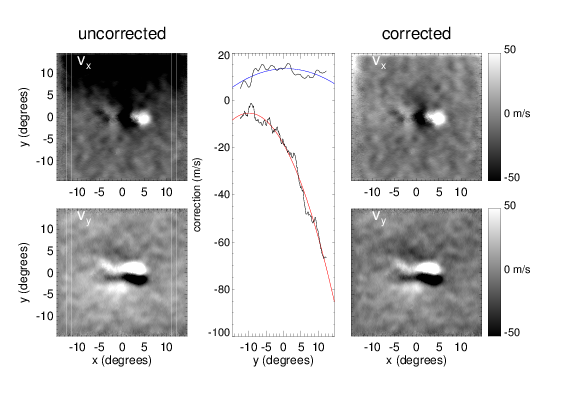}
\caption{
Demonstration of the removal of large-scale background flows 
from the flow components for group d (results for other groups are similar). 
The left two panels show the
raw $v_x$ (top) and $v_y$ (bottom) components, with ``quiet'' bands outlined
by vertical white lines on the edges of the frame. The values in the quiet regions are
averaged horizontally across the two bands and plotted against $y$ in the
center line plot. Polynomial fits to these data are overlaid for the $v_x$ (red)
component and $v_y$ (blue) component, respectively. The flow component maps on the
right show the residuals after these polynomials are subtracted from each column of the
uncorrected maps. The grey scales on the right showing speed in $\rm m\,s^{-1}$ apply to 
both the uncorrected and corrected maps.
}
\label{fig.flat}
\end{figure*}

Large-scale background flows, including differential rotation and
meridional flow, are removed from the ensemble-averaged flow components by fitting 
a quadratic polynomial in $y$ to the average of two vertical ``quiet-Sun'' 
strips of the component maps situated
near the east and west edges of the averaged frame (Figure~\ref{fig.flat}).
The assumption that the background flows vary only with $y$ is based on the near
alignment of this axis with the meridional direction in each of the Postel projections
contributing to the ensemble averages.
We note that this detrending removes any signal due to the departure of the
real solar (latitude-dependent) rotation from the tracked Carrington rate. Implications 
of this detrending for our results are further discussed in \S\ref{sec.discuss}.

\subsection{Ensemble Averages}\label{sec.eavgs}

\begin{figure*}
\begin{center}$
\begin{array}{cc}
\includegraphics[width=0.34\linewidth,clip]{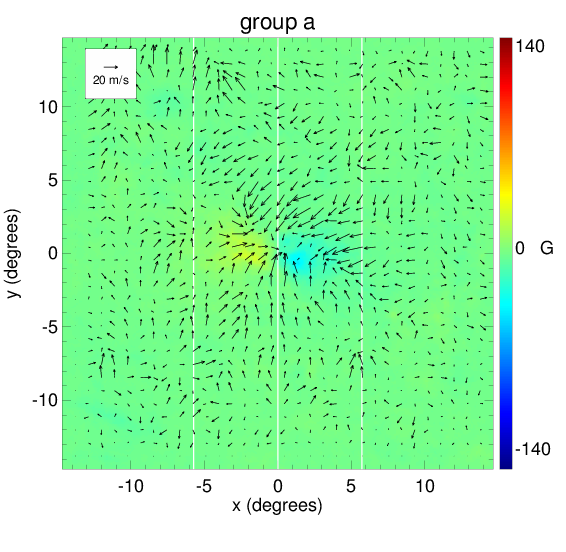}&
\includegraphics[width=0.34\linewidth,clip]{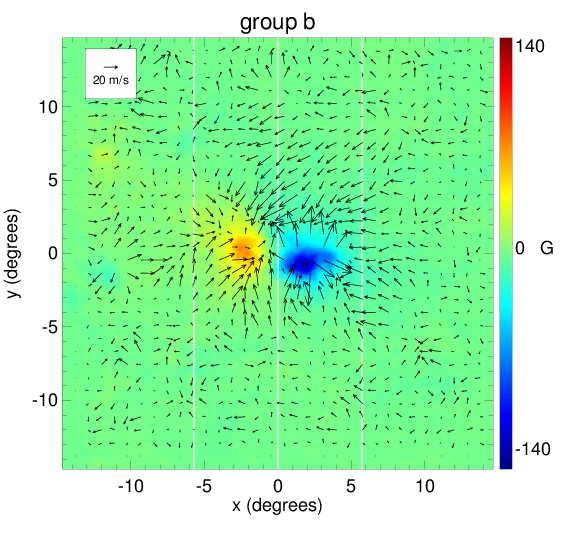}\\
\includegraphics[width=0.34\linewidth,clip]{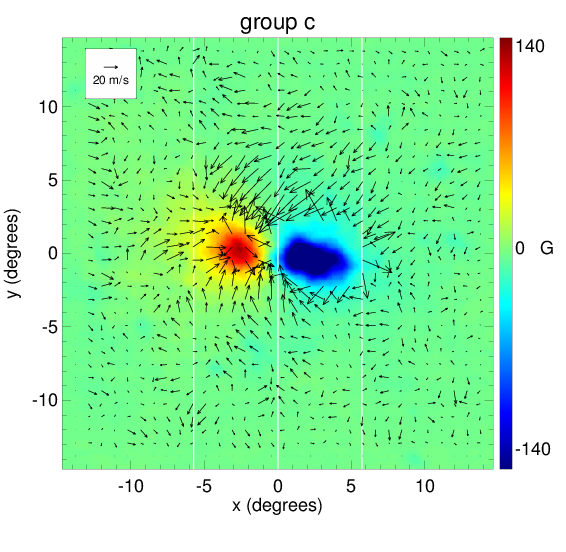}&
\includegraphics[width=0.34\linewidth,clip]{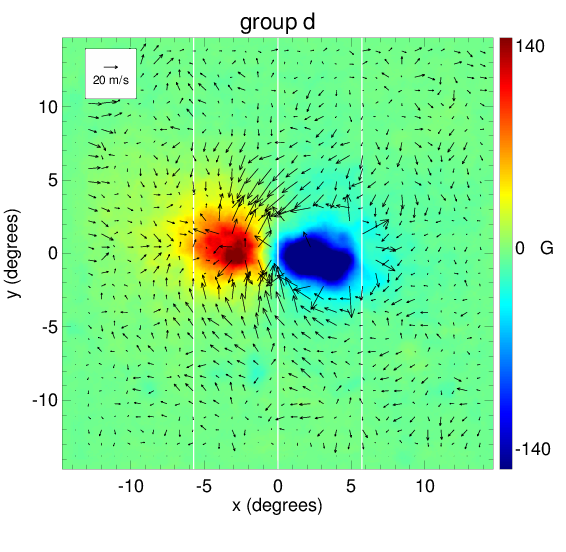}\\
\includegraphics[width=0.34\linewidth,clip]{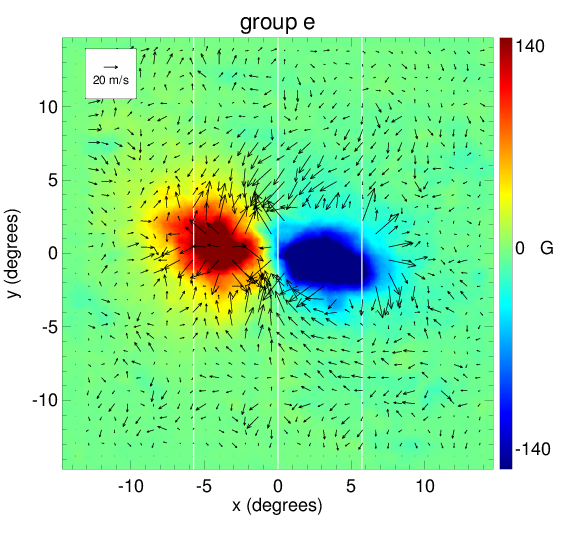}\\
\end{array}$
\end{center}
\caption{
Ensemble-averaged flows for each of the five flux groups as indicated. 
The background shows
the ensemble-averaged signed magnetic-flux
density.  As discussed in the text, a latitude-dependent
smoothly varying flow 
was assessed and subtracted from the flows shown here.
Converging flows are
prominent above and below the ARs.
Outflows from
sunspots are suppressed in the larger flux groups
(groups c,d, and e) for clarity.
Vertical white lines isolate the leading (right)
and trailing (left) regions of the active regions for further analysis 
(see text).
}
\label{fig.eavgs}
\end{figure*}

Figure~\ref{fig.eavgs} shows the
ensemble averaged flows for each flux group after detrending is applied.
For the purpose of clarity, strong diverging flows due to sunspots have been suppressed 
in Figure~\ref{fig.eavgs}. However, a version of this figure showing the complete flow fields
(Figure~\ref{fig.eavgs2}) is presented in Appendix~\ref{sec.eavgs2}.

Whereas comparable flow maps of individual active regions show root-mean-squared (RMS) 
fluctuations of $\sim 100$ m s$^{-1}$ due to supergranulation, 
we estimate a $\sim 5$ m s$^{-1}$ error in each flow component of the averages shown in 
Figure~\ref{fig.eavgs}.
This value is consistently derived from measurements of both: 1)
the RMS pixel-to-pixel fluctuations within selected regions of the maps 
and 2) the RMS fluctuations, at a given pixel, among subsamples of ARs in each flux group.
In \S\ref{sec.quant} we show quantitative comparisons of some aspects of these flow maps,
but some general findings from  these maps are notable. 
In particular, all flux groups show flows converging from nearby quiet Sun regions 
extending between 2 and 10 $^{\circ}$ from both the polar and equatorial sides of the AR.

Despite a wide variation of magnetic flux across the different
groups, both the amplitude and spatial extension of the inflows appear similar for
all groups.  Another distinct trend is that, for most flux groups,
the bulk of the flows in both of these flanking regions have a 
net eastward (retrograde) component. 
Both of these findings are discussed below.

\subsection{Flow Variations with AR Flux}\label{sec.quant}

To compare the flows around ARs among different flux groups, we average
the flow components over a modest range in longitude. Longitudinal averages are useful 
to examine the potential contribution of these AR-related flows to global meridional
or zonal (e.g.\ torsional oscillation) patterns and is discussed 
in \S\ref{sec.global}. Motivated by apparent asymmetries between
the two polarity regions, we average over 6$^{\circ}$ longitude ranges isolating 
either the leading and trailing
regimes as defined by the vertical white lines in Figure~\ref{fig.eavgs}.

\begin{figure*}
\includegraphics[width=0.48\linewidth]{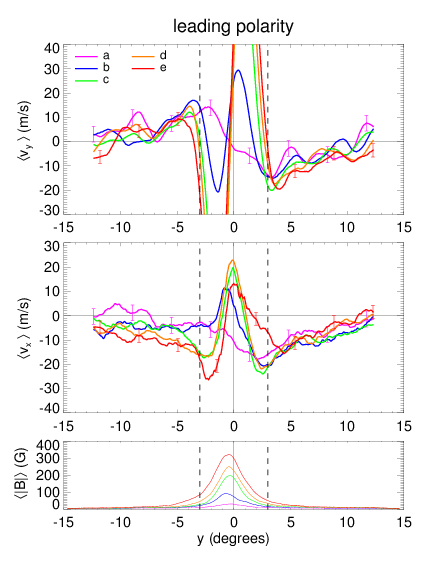}
\includegraphics[width=0.48\linewidth]{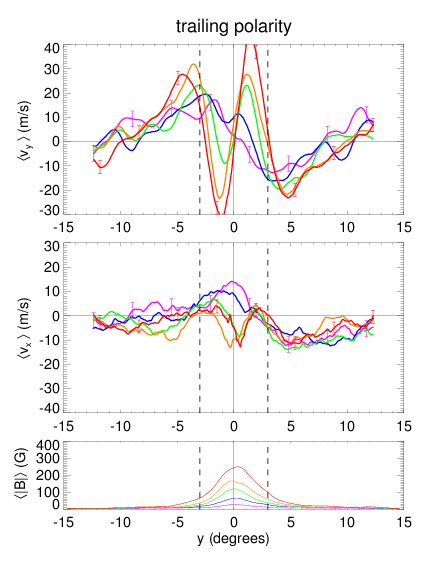}
\caption{
Averages over longitude of the group-averaged flow components, and magnetic flux density, 
across the leading and trailing polarity
zones. The panels on the left (right) show the results for the leading (trailing)
zones, with the top (middle) panels showing the averages of $v_y$ ($v_x$) respectively.
Each group is denoted by a different line color as indicated in the figure legend.
Only a few one-sigma error bars (denoting the error of the mean over the group) are shown
to avoid clutter and are typical of all of the errors. 
Averages of the unsigned magnetic flux density over each of the polarity zones are shown in the bottom 
panels, with the same colors as the flow averages. 
The vertical dashed lines bound the region within $\pm 3^\circ$ of the AR center to guide the 
relevant discussion in the text.
}
\label{fig.lavgfigs}
\end{figure*}

Further discussion of our findings separate the inferred flows into two regimes: those
confined to within 3$^{\circ}$ of the AR center latitude and those extending beyond this
distance. The reason for this distinction is the greater uncertainty in inferences in the former
group, based on our tests of helioseismic holography using MHD simulations of both sunspots
and quiet-Sun convection (Appendix~\ref{sec.valid}).  Those tests confirm that strong
deviations between inferred and true flows exist within the penumbral boundary of
simulated spots. In addition, while the moat flows (extending about 30 Mm from the 
simulated sunspot) are qualitatively
reproduced, the amplitudes appear to be systematically underestimated 
(Figure~\ref{fig.simoutflow}). Appropriate caveats should therefore be applied to the
results relevant to the flows near or within the ARs themselves, which we discuss first. 

Close to the AR centers 
(i.e.\ within the vertical dashed lines in Figure~\ref{fig.lavgfigs}) 
outflows are present 
in the meridional flow component $v_y$ from both polarities of most magnetic regions. 
Only the weakest ARs (group a) do not show a central divergence from the leading polarity, 
while all but  groups a and b show diverging flows from the trailing polarity. 
The outflows increase in magnitude with higher flux, which is likely an indication
of an increasing number of larger sunspots.

The zonal component $v_x$ within magnetic regions also shows variations with flux.
Within the leading polarity, most ARs (with the exception of group a) 
show prograde flows. This is consistent with prior inferences of prograde motion in ARs
\citep[e.g.][]{Zhao2004,Braun2004}. 
In the trailing polarity, the stronger regions (groups c through e) 
exhibit retrograde motions while the
weaker groups (a and b) show prograde flows. 
A simple picture at least roughly consistent with these trends is that,
with respect to the polarity boundary, one observes 
diverging flows from stronger regions but convergence in weaker regions.

Turning to the flows extending beyond 3$^{\circ}$ of the AR center latitude 
(i.e.\ outside of the vertical dashed lines in Figure~\ref{fig.lavgfigs}),
the predominant inference from the meridional components is the presence of 
inflows extending up to about 10$^{\circ}$ from the AR centers. These inflows
sometimes have magnitudes as strong as 20 or 30 m s$^{-1}$ close to the
AR, but are typically about 10 m s$^{-1}$ further away from the ARs.
From the meridional components alone it is apparent that the inflows are
stronger into the trailing polarity than the leading polarity.
The converging flows into the trailing polarity appear to increase in 
amplitude (with reasonable significance above formal errors) with increasing flux, 
but are invariant (within errors) with respect to flux into the leading polarity.
The inflows are mostly symmetric about the central latitude. 

Eastward (retrograde) flows are present in most regions, also with 
10 m s$^{-1}$ amplitudes and extending to 10$^{\circ}$ from the AR centers. 
Retrograde flows flank both the leading and trailing polarities,
suggesting that they are a feature distinct from, and not simply caused by,
the convergence towards the trailing polarity.
In general, the retrograde flows occur predominantly on the polar side.
This asymmetry is particularly notable above and below the trailing polarity in all groups. 
In the leading polarity, the retrograde flows increase with flux on the equatorial side,
such that the strongest regions show similar amplitudes on both sides.

A reasonable question is how systematic errors due to strong magnetic  
magnetic fields (as revealed, for example, in Appendix~\ref{sec.valid}) 
may influence these inferences 
given the averaging over many active regions with different morphologies.
For example, individual ARs may include sunspots or other strong field regions which are significantly
displaced by many degrees from the center of the averaged AR, and thus potentially compromise
measurements of flows at these corresponding locations. 
Appendix~\ref{sec.mask} shows the results of a direct test of this issue, whereby ensemble
averages are performed with spatial masks applied to exclude flows within strong magnetic fields.
These tests indicate a high robustness of flow inferences 3$^{\circ}$ away from the central
AR latitude.

\subsection{Contribution to Global Meridional and Zonal Flows}\label{sec.global}

Flows surrounding active regions may
contribute, even inadvertently, to assessments of the longitudinal averages
of global flow patterns including the differential rotation and meridional circulation.
Quantifying this contribution is particularly important in interpreting or modeling
the solar-cycle variations of these flows and, in turn, making long-term predictions about the
Sun. Measurements of 
the 11-year variability in both the
meridional circulation and the differential rotation (with the latter 
variability dominated by the ``torsional oscillation'') indicate 
amplitudes on the order of a few m s$^{-1}$ \citep[e.g.][]{Howe2009,Gizon2010b} 
and, for the most part, are spatially 
correlated with latitudes of solar activity. AR inflows and, likely the retrograde flows
which accompany them, have sufficient magnitude to contribute
to these global measurements. For example, just three ARs on the Sun at sufficiently
close latitudes would contribute about a 1 m s$^{-1}$ signal to a longitudinal average
of either the meridional or zonal flow, assuming each AR is characterized with a 10 m s$^{-1}$
flow spanning about 12$^{\circ}$ (the combined range of both polarity regions
as defined in this work). The ensemble averages presented above demonstrate that even weak ARs
(with fluxes on the order of a few $10^{21}$ Mx) could contribute detectable amounts to
these global flows.

\begin{figure}
\plotone{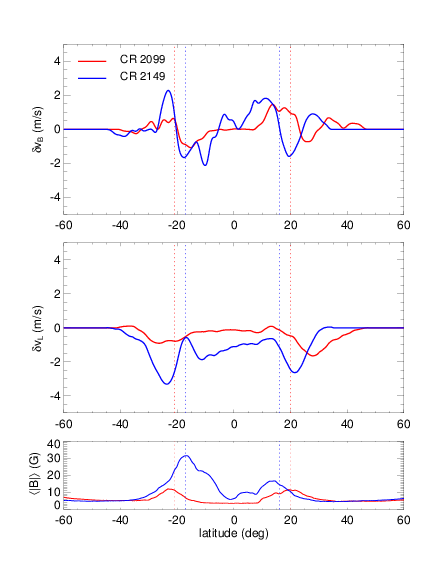}
\caption{
Hypothetical contributions to large-scale, longitude averaged, meridional (top panel) 
and zonal (middle panel) flows from active regions sampled in the 
synoptic maps of Carrington rotations 2099 and 2149
(i.e. close to solar minimum and maxima of cycle 24 respectively). A box-car smoothing 
with a width of 6$^{\circ}$ in latitude has been applied to the curves shown here.
The bottom panel shows the longitudinal average of the line-of-sight magnetic flux
density from HMI synoptic magnetograms. Vertical dotted lines in all panels 
indicate the centroid positions of the magnetic flux density as shown in the bottom panel.
}
\label{fig.global_contr}
\end{figure}

The large reduction of (largely supergranulation-dominated) noise in our averaged AR flow maps
makes feasible the use of these maps in making improved predictions on AR contributions
to global flows. We present a simple proof-of-concept of this, while leaving
a more detailed analysis to future work. This demonstration makes
use of the synoptic maps for Carrington rotations 2099 and 2149 
near solar minimum and maximum respectively (Figure~\ref{fig.cr2149} 
shows the magnetogram for CR 2149).
For each rotation, we identify the flux group of all of the ARs included in our survey
and remap the appropriate group-averaged flow components back to Carrington coordinates.
Longitudinal averages are then performed, with the results shown in 
Figure~\ref{fig.global_contr}. Smoothing (over a width of 6$^{\circ}$ in latitude) 
is applied to the curves shown in this plot, in order to simulate a more realistic assessment
(which, for example, might involve averages over multiple rotation periods). 
The use of nearby quiet-Sun regions to fit and remove background trends (\S\ref{sec.detrend})
implies that the net contributions shown in Figure~\ref{fig.global_contr} represent 
hypothetical perturbations to the mean (quiet-Sun) meridional and zonal flow components.

The predicted contributions ($\delta v_L, \delta v_B$) to the zonal and meridional 
flows respectively are shown in Figure~\ref{fig.global_contr}. They are both on 
the order of several m s$^{-1}$, which
confirms that the AR flows characterized in this survey are plausible detectable 
in global averages. It is noteworthy that, for the meridional component, the perturbations
to the mean meridional circulation take the form of converging zones centered (at least
approximately) on the mean latitudes of activity, which is 
qualitatively consistent with the observed modulation of meridional circulation observed near
solar maximum \citep[][]{Chou2001,Gizon2003,Zhao2004,Zhao2014}.

The results for the zonal components show predicted perturbations of similar magnitude
to the meridional components, but having a notably different variation with latitude.
In particular, the contributions are predominantly of one sign (corresponding to a net retrograde
motion) and have peaks
displaced significantly in latitude towards the poles relative to the mean AR latitudes.
This offset is apparently attributable to the polar/equatorward amplitude asymmetry in the 
retrograde flows noted earlier. For both rotations, these retrograde flows appear to 
dominate or at least cancel
out, the more spatially compact prograde motions observed in (primarily) the strongest ARs. 
We note the consistency of the residual retrograde flows observed in Figure~\ref{fig.global_contr}
with the general pattern the torsional oscillation, whereby active 
regions tend to center on the boundaries
between faster (in the polar direction) and slower (in the equatorial direction) flow bands.
Further interpretations of the results shown here need to be tempered by the intimidating number of
details specific to how the zonal-flow residuals, as well as the meridional flow, are obtained. 
Nevertheless, it seems unavoidable that active region flows provide 
a detectable contribution to global flow measurements.

\section{Discussion and Conclusions}\label{sec.discuss}

We have measured and compared active region flows, as determined using helioseismic holography,
and averaged over samples of about 1000 measurements within each of five groups of magnetic
flux. Principle results include the detection of both converging and retrograde motions extending
out to $\sim 10^{\circ}$ in latitude beyond the AR centers. Differences between flows associated
with the leading and trailing polarities, and among the difference flux groups are
noted in \S\ref{sec.quant}. 

Our measurements of the converging flows demonstrate the 
consistency of their general properties across a range of flux
exceeding an order of magnitude.
These inflows are similar to those inferred from
ensemble averages made with local correlation tracking methods 
\citep[][]{Loeptien2017}, including a preferential 
inflow to the trailing polarity. Our study uses a larger set of AR flow 
measurements, and includes ARs with weaker flux, than \cite{Loeptien2017}.
The converging flows we observe are also 
consistent with the inflow values of $20-30$ m s$^{-1}$ which \cite{Hindman2009}
found flowing into the ``periphery'' of ARs, as defined by a 50 G contour
in MDI magnetograms after smearing to a $2^{\circ}$ resolution.

Our ensemble averages show the presence of retrograde flows which 
straddle (primarily) the polar and (to a lesser degree) equatorial 
sides of ARs across the entire flux range studied. These appear to be 
heretofore unknown or, at least, unresolved in prior studies. However,  
the predominance of 10 m s$^{-1}$ retrograde flows towards the polar side of ARs is
plausibly consistent with the net cyclonic circulation of
$\sim 5$ m s$^{-1}$ found by \cite{Hindman2009} around AR peripheries. 
On the other hand, the retrograde flows do not readily appear in the ensemble
averaged flow map shown in Figure 8 of \cite{Loeptien2017}. This discrepancy
is not understood, but may result from differences in the background
trend removal or other details in the analyses. Applying both HH and 
local-correlation-tracking
methods to identical AR samples would be highly useful in exploring and understanding these
differences. 

Based on the estimates presented in 
\S\ref{sec.global}, the degree to which the retrograde flows may 
contribute to published measurements of the torsional oscillation 
is a fair question. It is worth emphasizing that the ensemble-averaged flows
shown in Figure~\ref{fig.eavgs} represent residuals after the 
subtraction of a latitude-dependent background contribution assessed
from quiet regions straddling the ARs. 
This procedure removes any real latitude-dependent 
zonal flow component which is invariant with longitude but departs from
the tracked Carrington rotation rate.
Therefore, the retrograde
flows presented here are spatially associated with the active regions, represent
a departure from the zonal flow present in the nearby quiet regions,  and
do not, for example, result from the choice of tracking rate. 
Further studies, similar to those carried out by \cite{Gonzalez-Hernandez2008}
and designed to isolate and compare flow contributions between quiet and active
regions, are critical to fully understanding the relation 
of the AR-specific flows presented here with longitudinal averaged global flows.

It would be useful to improve and expand the analysis presented here in order to
study the dependence of flows with other active-region properties, or to examine
their temporal variation as ARs evolve. Extending the analysis to regions 
with fluxes below $10^{21}$ Mx is also of interest.
Determining and validating the depth variation of the 
flows, particularly using helioseismic methods with high spatial resolution,
remain an important area for study. 
Inferring subsurface flows from travel-time measurements potentially compromised by the 
presence of strong magnetic fields remains a critical problem in local helioseismology.

\acknowledgements
Martin Woodard provided valuable comments on a draft of this work.
We are grateful to Charles Baldner and the rest of
Helioseismic and Magnetic Imager (HMI) team at Stanford
University for computing and providing the custom datasets for
this study. 
This work is supported by the Solar Terrestrial program of the National Science 
Foundation (grant AGS-1623844) and by the
NASA Heliophysics Division through its 
Heliophysics Supporting Research (grant 80NSSC18K0066) and
Guest Investigator (grant 80NSSC18K0068) programs.
Resources supporting this work were provided by the NASA High-End Computing 
(HEC) Program through the NASA Advanced Supercomputing (NAS) Division at Ames Research Center.
SDO data is provided courtesy of NASA/SDO and the AIA, EVE, and HMI science
teams.

\appendix

\section{Validation of Calibrated Flows using MHD Simulations}\label{sec.valid}

Validating methods used to infer flows, particularly in the presence 
of strong magnetic fields typical of active regions, is critical to correctly 
interpreting the results.  Uncertainties and discrepancies in helioseismic 
inferences about sunspots are well known \citep[e.g.][]{Gizon2009,Moradi2010,Braun2012}.
Effects which contribute to these uncertainties include, but are not
limited to, 1) the likely presence of strong near-surface magnetic and thermal 
perturbations in sunspots which are inconsistent with the Born 
approximation or other enabling assumptions 
\citep[e.g.][]{Braun2006,Couvidat2007,Crouch2010,Braun2012}, 
2) wave phenomena including absorption and mode
conversion \citep[e.g.][]{Woodard1997,Crouch2005b,Schunker2008}, 
3) consequences of magnetically suppressed p-mode amplitudes
\citep{Rajaguru2006}, and 4) the distortion of the spectral line used to assess Doppler
shifts in strong fields as well as instrumental and calibration limitations
\citep[e.g.][]{Wachter2006,Rajaguru2007}. Some of these phenomena
have been investigated in the context of thermal structure modeling but less
is known about their potential influence on flow measurements.

Agreement between flows inferred from different helioseismic methods \citep[][]{Hindman2004}, and 
between helioseismic and non-helioseismic methods \citep[][]{Liu2013,Birch2016,Jain2016}, provide
some confidence in the procedures and the results obtained.
On the other hand, these comparisons are potentially compromised by the lack of knowledge of
the actual flow fields present. This problem is resolved by the use of artificial data, such
as provided in realistic magnetohydrodynamic (MHD) simulations. Here, we use two
publicly available\footnote{\url{http://download.hao.ucar.edu/pub/rempel/sunspot\_models}} 
target simulations. The first of these is of quiet-Sun convection in the presence
of a small-scale dynamo \citep{Rempel2014b} which has been used to validate
forward and inverse methods in both time-distance helioseismology \citep{DeGrave2014a} and
helioseismic holography \citep{DeGrave2018}. The simulation covers a 98.304 $\times$ 
98.304 $\times$ 18.432 Mm domain, computed with 64 $\times$ 64 $\times$ 32 km resolution,
a total duration of 30 hr, and a time step of 0.9 sec. As in \citet{DeGrave2018},
we employ the first $15$~hr of artificial Dopplergrams from this run. 
The second simulation is a sunspot with a magnetic flux of approximately 
6 $\times 10^{21}$ Mx and contained in an identical domain, but computed at 
a 48 $\times$ 48 $\times$ 24 km resolution and a time step of 0.45 sec. The simulation 
shows a sunspot with a penumbra and Evershed flow \citep{Rempel2015}. This run
was carried out for 100 hr, of which we use a 15 hr interval starting 57.5 hr into
the run, by which time a realistic moat flow has developed.

\begin{figure*}
\includegraphics[width=0.45\linewidth]{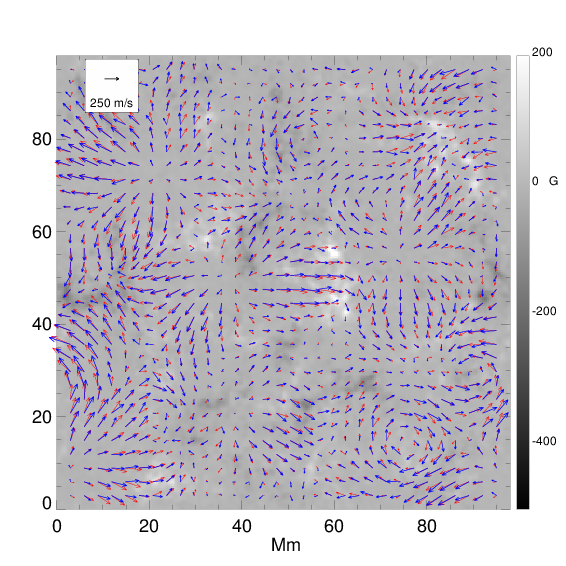}
\includegraphics[width=0.45\linewidth]{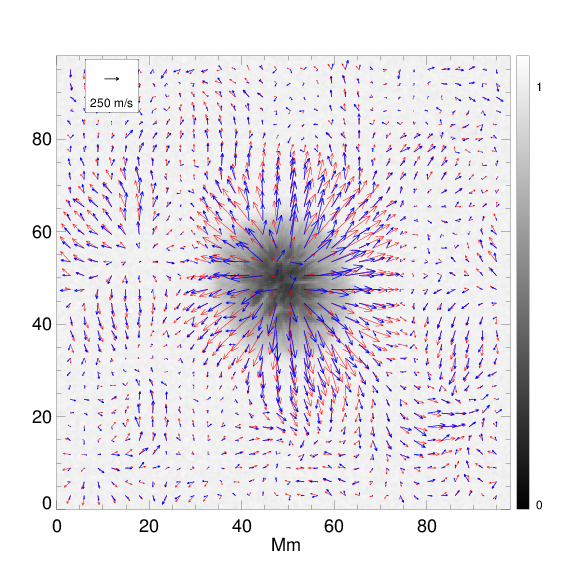}
\caption{
Comparisons of true flows (red arrows) with calibrated-HH flows (blue arrows) using artificial
data extracted from two numerical simulations (see text). The left panel shows the results
for a quiet-Sun convective simulation, with the background indicating the time average of
the vertical magnetic flux density. Temporal averaging and spatial smearing 
of the true flows are performed
as described in the text. The flows are comparable to those present in
solar supergranulation in both size and magnitude. A correlation coefficient of 0.92 is found
between the individual flow components ($v_x$, $v_y$) of the true and inferred flows.
The right panel shows the results for the sunspot simulation, with the background showing 
a snapshot of photospheric intensity normalized to the quiet photosphere. 
While there is qualitative agreement between the true and inferred moat flows (with a correlation
coefficient of 0.85), the inferred outflows
inside (outside) the penumbra appear overestimated (underestimated) compared to the actual flows.
}
\label{fig.vectsims}
\end{figure*}

The background atmosphere in both simulations has a vertical stratification 
consistent with Model S, for which we apply lateral-vantage holography with unmodified Greens 
functions \citep[see, for example][]{DeGrave2018} at a focus depth of 3 Mm. For comparison 
with the measured calibrated flows, we use time-averages of the true flows present in the 
simulation as sampled at fixed intervals (which are 83.3 and 30 min for the quiet-Sun convective and
sunspot simulations respectively). Consistent with the working assumptions for the calibrated-HH
method (\S\ref{sec.ttmeas}), we extract only the horizontal flow components and apply
a horizontal smoothing with a two-dimensional Gaussian function with a FWHM of
14 Mm in the direction along the component axis, and 9 Mm in the perpendicular direction. 
This smearing is comparable to the horizontal extent of the appropriate sensitivity functions
as illustrated by \citet{Braun2007} and \citet{DeGrave2018} and thus
removes small-scale noise.
The true flows are also averaged in depth with
a weighting given by the function shown in Figure~\ref{fig.kernz}. Figure~\ref{fig.vectsims} 
shows comparisons of the calibrated flows with the temporally averaged and spatially smeared 
true flows for both simulations. 

\begin{figure*}
\plotone{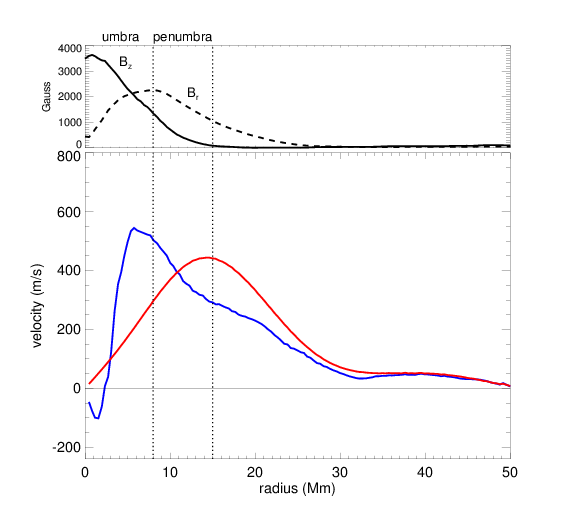}
\caption{
The top panel shows azimuthal averages of the vertical ($B_z$) and radial ($B_r$) component of 
the photospheric magnetic flux density from a time-average of the sunspot simulation.
The bottom panel shows azimuthal averages of radial (positive $=$ outward) flow components.
The radial component (of either the flux density or flow) is defined as the projection
of the horizontal component along a radius extending from the center of the sunspot. 
The red curve shows the result for the true flows (after time-averaging and spatially smearing). 
The blue curve shows the result for the calibrated-HH flows. Discrepancies between the two are
most obvious within the spot but also extend notably beyond the penumbra.
}
\label{fig.simoutflow}
\end{figure*}

The results show good agreement in the quiet-Sun convective case, as well as at least qualitative
agreement for the sunspot flows. 
For the quiet and sunspot simulations we find a correlation coefficient of 0.92 and 0.85,
respectively, between the individual flow components ($v_x$, $v_y$) of the true and inferred flows.
However, a closer examination shows that the inferred Evershed flows (within the penumbra) 
using the calibrated-HH method appear to be overestimated while 
most of the moat flows (extending
beyond the penumbra) are underestimated. This is best seen in Figure~\ref{fig.simoutflow} which shows
the azimuthal averages of the radial flow component (with respect to the sunspot center).

\begin{figure*}
\plotone{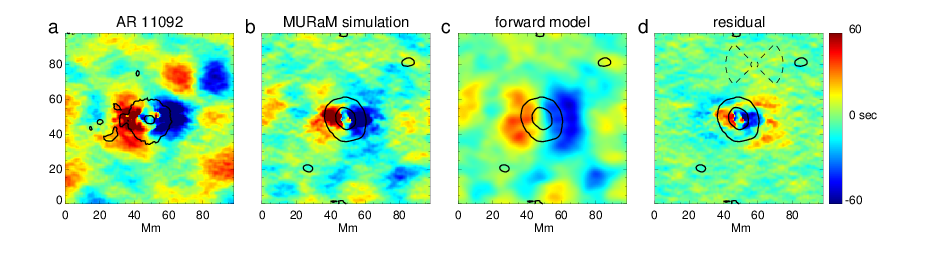}
\caption{
Comparison of west-minus-east (WE) travel-time differences. 
Panel (a) shows the WE travel-time difference for one the sunspot regions used in this study
(AR 11092). The area is cropped to the same dimension of the WE travel-time difference
measured in the sunspot simulation of \citet{Rempel2015} which is shown in panel (b). 
Panel (c) shows the predicted travel-time difference (or ``forward model'') obtained 
by a convolution 
the sensitivity function with a time-average of the three-dimensional flows present in
the simulation. Panel (d) shows the residual signal after subtracting the forward model
from the measurement. Black contours indicate time-averaged 
vertical magnetic flux density values of 300 and 2000 G. In panel (d) the dashed lines 
show the geometry of the east and west pupils used in the measurement.
}
\label{fig.artifact}
\end{figure*}

To explore these discrepancies further, we employ the sensitivity functions derived
in prior work \citep{DeGrave2018} to examine differences between the actual travel-time
difference measurements with those predicted under the assumption of the Born approximation.
Such a comparison removes potential issues arising from the
assumptions specific to the calibrated HH method. 
A detailed study of the quiet-Sun simulation, in the
context of the forward-modeling problem, was already performed by \cite{DeGrave2018}.
They found that a comparison of all lateral-vantage HH
measurements (including the focus depth of 3 Mm employed here) with the forward prediction
from the sensitivity functions showed excellent agreement within the realization
noise. A comparison of results for the sunspot simulation, however, gives a notably different
picture (e.g.\ Figure~\ref{fig.artifact}). Specifically, the measured travel-time differences
show the presence of an artifact which is revealed after the predicted model travel-time
differences are subtracted from the measurements. This artifact takes the form of a spurious
outflow within the penumbra in addition to another spurious inflow extending about 15 Mm
beyond the penumbra. 
The cause of this artifact is unknown, but apparently it represents a
failure of the sensitivity function to fully account for the physics of wave propagation
in the presence of strong magnetic flux (examples of relevant issues are
listed in \S\ref{sec.valid}). We note that this artifact has implications beyond the
use of the calibrated-HH method and will presumably adversely affect any inverse or forward
modeling based on standard (Born approximation-based) methodology.
Comparisons using different lateral-vantage HH measurements confirm that 
the spatial extent of the artifacts are apparently related to the geometry of the pupils used.
Notably, the maximum amplitude of the artifacts appear in locations where either the
umbra or penumbra fill one of the pupil quadrants.

\section{Ensemble Averages Including Sunspot Flows}\label{sec.eavgs2}

\begin{figure*}
\begin{center}$
\begin{array}{cc}
\includegraphics[width=0.34\linewidth,clip]{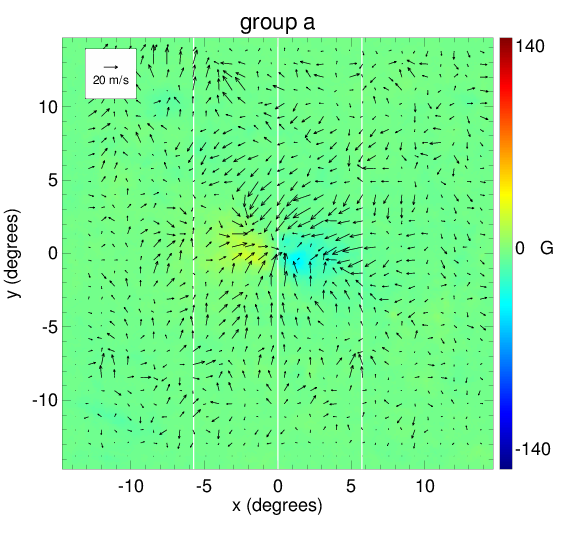}&
\includegraphics[width=0.34\linewidth,clip]{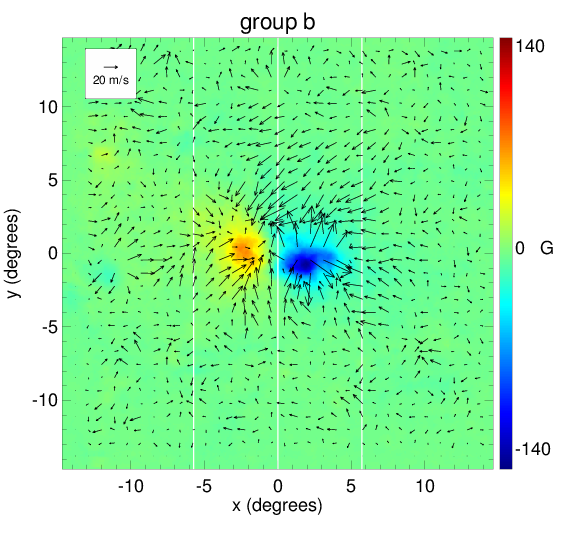}\\
\includegraphics[width=0.34\linewidth,clip]{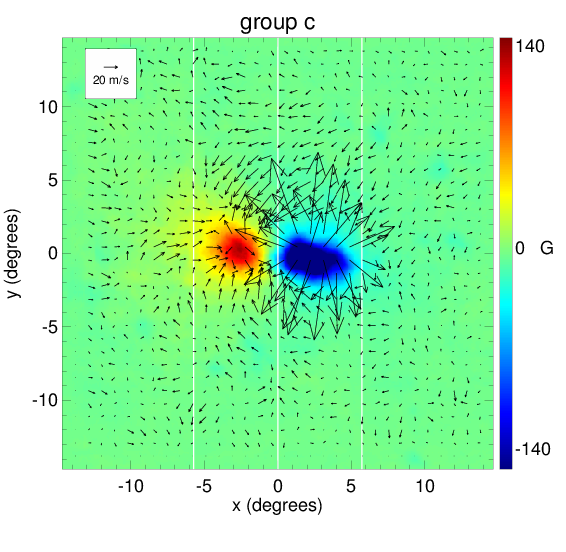}&
\includegraphics[width=0.34\linewidth,clip]{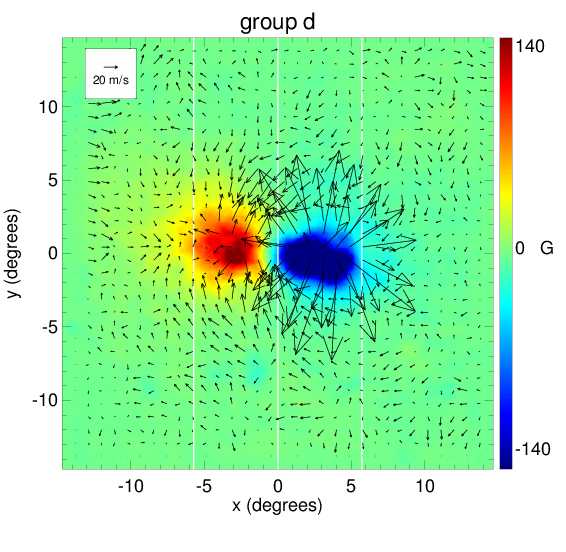}\\
\includegraphics[width=0.34\linewidth,clip]{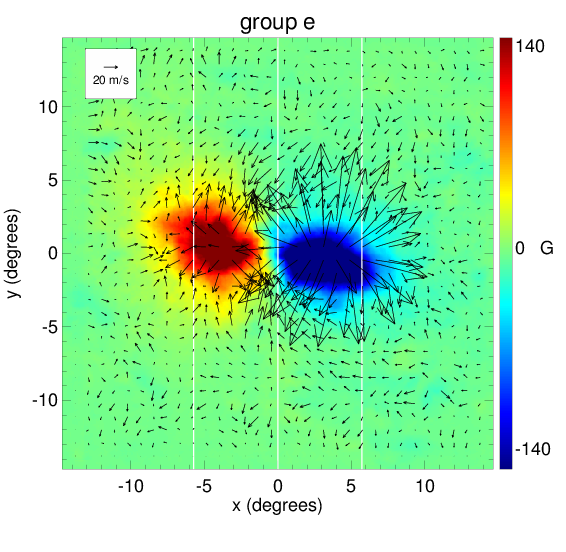}\\
\end{array}$
\end{center}
\caption{
Ensemble-averaged flow fields for each flux group, as in Figure~\ref{fig.eavgs}, but without the suppression of
the strong, mostly diverging, flows due to the contribution of sunspot
moats in the averaging. 
}
\label{fig.eavgs2}
\end{figure*}

Figure~\ref{fig.eavgs2} shows a version of Figure~\ref{fig.eavgs}
without the suppression of the strong (mostly diverging) flows observed
in the centers of the ARS. These diverging flows arise from the
contribution of moat flows around sunspots in each flux group. 

\section{Magnetic Masking Test}\label{sec.mask}

\begin{figure*}
\plotone{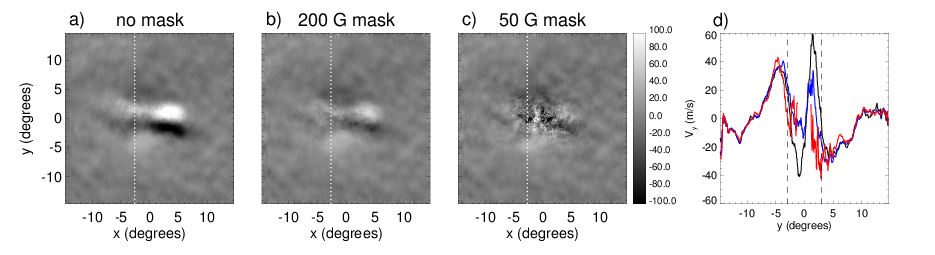}
\caption{
Results from masking tests designed to reveal the dependence on (and possible contamination from)
the inclusion of strong magnetic regions in the ensemble averaging.
Panel (a) shows the poleward flow component $v_y$, averaged over the ARs in group e,
and after detrending (\S\ref{sec.detrend}). Panel (b) shows the same average, but
after masking out flows within pixels of the individual flow maps with 
flux density greater than 200 G.
Panel (c) shows the average using a magnetic threshold of 50 G. 
Panel (d) shows a comparison of vertical cuts along the dotted lines shown in
panels (a) - (c). The black curve indicates the results for the unmasked flows,
while the blue (red) curve shows the results using the 200 (50) G masks. 
The vertical dashed lines bound the region within $\pm 3^\circ$ of the AR center. 
Within the AR center the strong flow signatures are reduced 
using the 200 G mask and become noisy with the 50 G mask.  
A portion of the 50 G measurements in panel (d) close to the AR is omitted 
to avoid obscuring the plot with noise. 
The main result of this test is that the 
more extended converging signatures remain robust with respect 
to the masking. 
}
\label{fig.mask}
\end{figure*}

Given the results of our validation studies with numerical simulations (Appendix~\ref{sec.valid})
we investigate how our flow measurements may be compromised by artifacts caused by
regions of magnetic flux typical of sunspots. To assess the degree of potential contamination of our results,
we perform ensemble averages of the flow components as in \S\ref{sec.eavgs} but set 
the flows in the individual maps to zero in regions above a given magnetic 
flux density. This data masking is not meant as a ``correction,'' 
but is rather a test of where and how much results change between the 
masked and unmasked averaging. Control tests with magnetic masks are commonly 
used in helioseismology \citep[e.g.][]{Zhao2003,Korzennik2006,Gonzalez-Hernandez2008,Liang2015a}.
Figure~\ref{fig.mask} shows the results for the strongest flux group, representing the worse-case
scenario, and using masks with a threshold flux densities of 200 and 50 G. The masks 
are constructed
using a potential-field extrapolation of the total field from the line-of-sight 
(time-interval averaged) magnetograms as described by \citet{Braun2016}. Examples of
the masks as applied to an active region are shown in Figure 1 of \citet{Braun2016}.
The 200 G mask removes flows within sunspot umbra and penumbra which are identified as the
most problematic regions in Appendix~\ref{sec.valid}. The 50 G mask removes weaker fields
around the ARs and provides even stronger constraints on potential contamination.
Results show little change in inferences of $v_y$ between masked and unmasked averaging  
for distances beyond about 3$^{\circ}$ from the AR center. 
Within 3$^{\circ}$ the flow signals are reduced using the 200 G threshold, 
apparently due to the exclusion of Evershed and other sunspot flows from the average. 
For the 50 G threshold, 
the averaged flows within the AR become dominated by noise. This 
is likely due to poorer statistics (i.e.\ fewer non-zero pixels) in the averaging.
The results, however, appear to establish a relatively high confidence in the more extended 
inflows and retrograde motions, which was the primary purpose of this test.

\bibliographystyle{/export/home/dbraun/Macros/apj}
\bibliography{/export/home/dbraun/Macros/db}

\end{document}